\documentclass[twocolumn,superscriptaddress,showpacs,preprintnumbers]{revtex4}
\usepackage{amssymb}
\usepackage{amsmath}
\usepackage{graphicx}
\usepackage{dcolumn}
\usepackage{verbatim}
\usepackage{bm}

\input{tcilatex}

\begin{document}

\title{Mixed-state fidelity and quantum criticality at finite temperature}
\author{Paolo Zanardi}
\affiliation{Institute for Scientific Interchange, Villa Gualino, Viale Settimio Severo
65, I-10133 Torino, Italy }
\affiliation{Department of Physics and Astronomy, University of Southern California Los
Angeles, CA 90089-0484 (USA) }
\author{H. T. Quan}
\affiliation{Institute of Theoretical Physics, Chinese Academy of Sciences, Beijing,
100080, China}
\author{Xiaoguang Wang}
\affiliation{Zhejiang Institute of Modern Physics, Department of Physics, Zhejiang
University, Hanzhou 310027, China.}
\author{C. P. Sun}
\affiliation{Institute of Theoretical Physics, Chinese Academy of Sciences, Beijing,
100080, China}
\date{\today}

\begin{abstract}
We extend to finite temperature the fidelity approach to quantum phase
transitions (QPTs). This is done by resorting to the notion of mixed-state
fidelity that allows one to compare two density matrices corresponding to
two different thermal states. By exploiting the same concept we also propose
a finite-temperature generalization of the Loschmidt echo. Explicit
analytical expressions of these quantities are given for a class of
quasi-free fermionic Hamiltonians. A numerical analysis is performed as well
showing that the associated QPTs show their signatures in a finite range of
temperatures.
\end{abstract}

\pacs{03.65.Ud,05.70.Jk,05.45.Mt}
\maketitle

\section{introduction}

Even, small and smooth changes in parameters governing the dynamics of a
physical system can, in some circumstances, result in a dramatic change of
physical properties of the system itself. The traditional approach to these
so-called critical phenomena is based on the notions of order parameter,
correlation functions, symmetry breaking, and a general formulation in the
framework of the Landau-Ginzuburg picture and renormaliztion group~\cite%
{huang}. A \emph{phase transition} can be triggered e.g., by a change of
temperature of the system or, at zero temperature, by a change of some of
the coupling constants ( e.g., external fields) defining the system's
Hamiltonian. In the first case one says that the transition is driven by
thermal fluctuations, and the transition is referred to as \emph{classical},
whereas in the second quantum fluctuations are held responsible for the
transition and this latter is referred to as a \emph{quantum phase transition%
} (QPT)~\cite{sachdev}.

In the last few years a big deal of interest has grown about the possibility
of studying QPTs by means of ideas and tools borrowed from the new born
field of Quantum Information Science \cite{qis}. In this novel approach the
key concept involved is quantum entanglement (or genuinely quantum
correlations) and the idea is that quantum criticality can be suitably
characterized in terms of the behavior of different entanglement measures
\cite{qpt-qis}. Though the picture is still rapidly moving it is by now
clear that, whereas there are no doubts that entanglement is indeed a
valuable conceptual tool to analyze QPTs, one has, case by case, identify
what is the entanglement measure e.g., block-entanglement or concurrence
most suited to extract the relevant information.

More recently an approach to QPTs based on another quantum information
notion i.e., \emph{quantum fidelity}, has been put forward~\cite%
{za-pa,za-co-gio}. The idea behind this novel approach is quite simple: the
dramatic chance of the structure of the ground state occurring at the
critical points can be fruitfully studied by analyzing their degree of
distinguishability. Since this latter quantity is related to the overlap
i.e., scalar product, between two different ground states, the fidelity
approach is basically nothing but than a \emph{metric} one; the key
ingredient being provided by the state-space distance between states
corresponding to two slightly different sets of coupling constants. The
expectation is that at the critical points a small change of the control
parameters should result in an enhanced modification of the state structure
and this in turn should be detected by a greater state-space distance i.e.,
statistical distinguishabiliy, between the associated quantum states. In
spite of its apparent naivety, this metric-based approach turns out to be
able to provide an effective way to obtain, qualitative as well as
quantitative, information about the zero-temperature phase diagram of a
large class of non trivial quantum systems i.e., quasi-free fermionic
systems \cite{za-co-gio} and matrix-product states \cite{co-ion-za}. The
conceptual appealing of the fidelity approach for detecting boundaries
between different phases lie in its universal geometrical as well as
information-theoretic nature; in principle no apriori knowledge of the
symmetry-breaking mechanism and of the associated order parameters is
required.

In this paper we are going to extend the fidelity approach to \emph{finite
temperature}. This generalization can be achieved by exploiting the
mixed-state fidelity introduced by Uhlmann~\cite{Uhlmann} and related to the
statistical distance between two density operators (Bures distance). This
finite temperature extension of the fidelity approach is a non-trivial
technical step necessary to analyze the signatures of QPTs at non zero
temperature and, more in general, to investigate the potential usefulness of
the fidelity notion in the study of classical i.e., temperature-driven phase
transitions \cite{gio}. In the following we will focus on the first task.
Moreover, we will provide an explicit finite-temperature extension of
another concept that has been recently used in the context of QPTs and
inspired the whole fidelity programme: the Loschmidt echo~\cite%
{zhong-guo,cuc}.

The paper is organized as follows: In Sec. II, we analytically calculate the
mixed-state fidelity of thermal states and a numerical analysis is used to
demonstrate the signatures of QPTs at finite temperature. In Sec. III, we
calculate the Loschmidt echo through a similar procedure. Sec. IV contains
the conclusion.

\section{Mixed-state fidelity of thermal states}

Let us consider the set of (mixed) quantum states $\mathcal{S}(\mathcal{H}%
):=\{\rho \in \mathcal{L}(\mathcal{H})\,/\,\rho \geq 0,\mathrm{{tr}\rho =1\}.%
}$ The mixed-state fidelity is given by \cite{Uhlmann}
\begin{equation}
F(\rho _{0},\rho _{1}):=\mathrm{{tr}\sqrt{\rho _{1}^{\frac{1}{2}}\rho
_{0}\rho _{1}^{\frac{1}{2}}}.}  \label{mix-fid}
\end{equation}%
This quantity measures the degree of distinguishability between the two
quantum states $\rho _{0}$ and $\rho _{1}.$ The fidelity is related to the
statistical Bures distance: $D(\rho _{0},\rho _{1})=\sqrt{2(1-F)}.$ We will
use Eq. (\ref{mix-fid}) to compare two different thermal states%
\begin{eqnarray}
\rho _{\alpha } &=&Z_{\alpha }^{-1}\exp (-\beta _{\alpha }H_{\alpha }),\
\notag \\
Z_{\alpha } &:&=\mathrm{tr}\exp (-\beta _{\alpha }H_{\alpha }),(\alpha
=0,1).
\end{eqnarray}%
We define
\begin{equation}
\mathcal{F}_{H_{0},H_{1}}(\beta _{0},\beta _{1}):=F\left(
Z_{0}^{-1}e^{-\beta _{0}H_{0}},Z_{1}^{-1}e^{-\beta _{1}H_{1}}\right) .
\end{equation}%
In particular one can consider the cases

\begin{itemize}
\item {(i)} $H_\alpha=H(\lambda_\alpha), \lambda_1=\lambda_0+\delta\lambda;
\beta_0=\beta_1=\beta.$

\item {(ii)} $H_{\alpha }=H(\beta _{\alpha }),\beta _{1}=\beta _{0}+\delta
\beta .$
\end{itemize}

The first case (i) is useful to study the finite-temperature signatures of a
QPT occurring at some points, say $\lambda _{c},$ in the parameter space.
The second case (ii) is considered for studying temperature driven phase
transitions. Here the $\beta $ dependence of the Hamiltonian can be the
result of the presence of e.g., a chemical potential term (grand canonical
ensemble) or of self-consistently determined coupling e.g., a mean-field BCS
pairing term. In this paper we will focus on case (i), while the use of
mixed-state fidelity for analyzing temperature driven PTs will be addressed
in a forthcoming work.

\begin{figure}[tbp]
\begin{center}
\includegraphics[bb=80 325 507 667, width=8cm, clip]{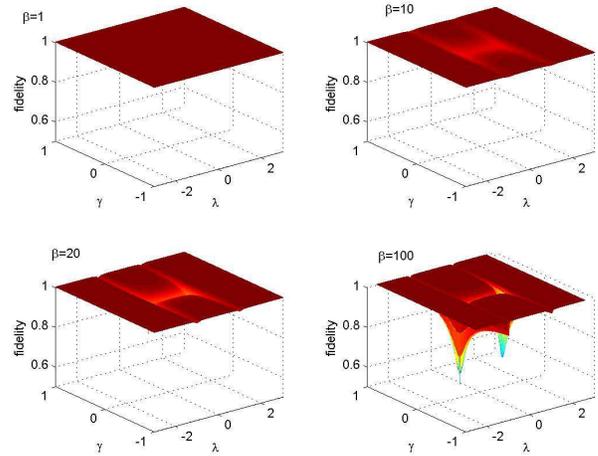}

\end{center}
\caption{(Color online) Mixed-state fidelity of finite-temperature thermal
state of the XY model. The fidelity is a function of $\protect\lambda $ and $%
\protect\gamma $. Here, we choose the spin number $N=200$ and the parameter
perturbation $\protect\delta \protect\lambda =\protect\delta \protect\gamma %
=10^{-2}$. The temperatures of system in the above four figures are chosen
to be $\protect\beta =1, \protect\beta =10, \protect\beta =20$, and $\protect%
\beta =100$, respectively. At low temperatures, e.g., $\protect\beta =100$,
the mixed-state fidelity clearly shows the signature of QPT which occurs at
absolute zero temperature. When the temperature increases, the decay of
fidelity becomes less sharp and finally the signature of QPT disappears due
to the thermal excitation.}
\end{figure}

\begin{figure}[tbp]
\begin{center}
\includegraphics[bb=37 191 570 612, width=7cm, clip]{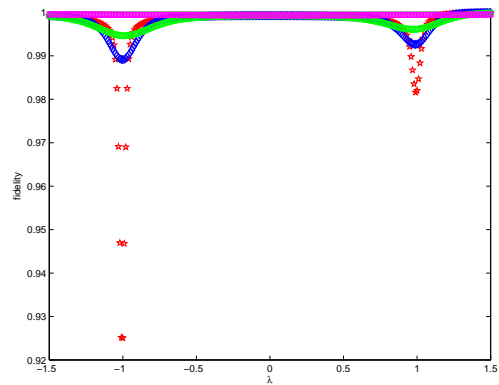}
\end{center}
\caption{(Color online) A cross section of Fig. 1 with $\protect\gamma=1$,
i.e., mixed state fidelity of finite-temperature thermal state of transverse
Ising model. Here, the same as that in Fig. 1, the spin number is chosen to
be $N=200$ and the parameter perturbation $\protect\delta \protect\lambda =%
\protect\delta \protect\gamma =10^{-2}$. Four curves represent four
different temperatures: $\protect\beta =1, \protect\beta =10, \protect\beta %
=20$, and $\protect\beta =100$ respectively. The decay of fidelity, though
becomes less sharp with the increase of the temperature, well indicates the
critical point of the transverse Ising model.}
\end{figure}

\subsection{Commuting Hamiltonians}

Let us start by considering the simple case where the two Hamiltonians $%
H_{0} $ and $H_{1}$ commute; then
\begin{equation}
\rho _{1}^{\frac{1}{2}}\rho _{0}\rho _{1}^{\frac{1}{2}}=\rho _{1}\rho
_{0}=(Z_{0}Z_{1})^{-1}\exp (-\beta _{0}H_{0}+\beta _{1}H_{1}).
\end{equation}%
Therefore by using the definition (\ref{mix-fid})
\begin{eqnarray}
&&F(\rho _{0},\rho _{1})  \notag \\
&=&(Z_{0}Z_{1})^{-1/2}\mathrm{tr}\exp \left[ -(\beta _{0}H_{0}+\beta
_{1}H_{1})/2\right]  \notag \\
&=&(Z_{0}Z_{1})^{-1/2}\sum_{n}\exp (-(\beta _{0}E_{n}^{0}+\beta
_{1}E_{n}^{1})/2),
\end{eqnarray}%
where $H_{\alpha }|\Psi _{n}\rangle =E_{n}^{\alpha }|\Psi _{n}\rangle
(\alpha =0,1).$ In particular if moreover $H_{0}=H_{1}$ one immediately
finds that fidelity of thermal states can be expressed entirely in terms of
partition functions
\begin{equation}
\mathcal{F}(\beta _{0},\beta _{1})=\frac{Z(\frac{\beta _{0}+\beta _{1}}{2})}{%
\sqrt{Z(\beta _{0})Z(\beta _{1})}}.  \label{Z}
\end{equation}%
We note in passing that this relation seems to suggest the possibility of
making a direct connection between the metric/statistical notion of fidelity
and purely thermodynamical quantities as well as the viability of the
fidelity approach even for classical systems \cite{gio}. We would like also
to observe that Eq (\ref{Z}) also gives the (pure-state) fidelity between
the pure quantum states that one associates to the thermal states $\rho_0$
and $\rho_1$ through the kind of classical-quantum correspondence discussed
in \cite{frank} [see Eq (1) there].

To further exemplify this commuting case let us consider diagonal fermionic
Hamiltonians $H_{\alpha }=\sum_{k}\epsilon _{k}^{\alpha }c_{k}^{\dagger
}c_{k}.$ One has
\begin{eqnarray*}
\rho _{\alpha }(\beta ) &=&Z_{\alpha }^{-1}\prod_{k}e^{-\beta _{\alpha
}\epsilon _{k}^{\alpha }c_{k}^{\dagger }c_{k}} \\
&=&\prod_{k}(1+e^{-\beta _{\alpha }\epsilon _{k}^{\alpha }})^{-1}\left[
1+(e^{-\beta _{\alpha }\epsilon _{k}^{\alpha }}-1)c_{k}^{\dagger }c_{k}%
\right] .
\end{eqnarray*}%
From this one finds
\begin{equation}
\mathcal{F}_{H_{0},H_{1}}(\beta _{0},\beta _{1})=\prod_{k}\frac{1+e^{-(\beta
_{0}\epsilon _{k}^{0}+\beta _{1}\epsilon _{k}^{1})/2}}{\sqrt{(1+e^{-\beta
\epsilon _{k}^{0}})(1+e^{-\beta \epsilon _{k}^{0})})}}.  \label{fid-diag}
\end{equation}%
In the zero-temperature limit $\beta \rightarrow \infty ,$ from the above
equation, it is easy to check that $\forall k,\epsilon _{k}^{0}\epsilon
_{k}^{1}\geq 0\Rightarrow \mathcal{F}_{H_{0},H_{1}}(\infty ,\infty )=1,$
wheras if $\exists k,\epsilon _{k}^{0}\epsilon _{k}^{1}<0\Rightarrow
\mathcal{F}_{H_{0},H_{1}}(\infty ,\infty )=0.$ This situation is the one
that encounters in the $XY$-model analysis of \cite{za-pa} in the critical
line with anisotropic parameter $\gamma =0.$ There indeed when the magnetic
field $\lambda $ (which plays the role of a chemical potential in the
fermionic picture) is changed in the range $[-1,1]$ one has exactly that one
of the single particle eigenvalues changes sign and this results in a
vanishing fidelity \cite{za-pa}.

\subsection{A more general case}

Now we move to consider a more general case directly relevant to the XY and
mean-field BCS-like models. The Hamiltonian is given by
\begin{equation}
H^{\alpha }=\sum_{k}H_{k}^{\alpha }=\sum_{k}\epsilon _{k}^{\alpha
}(n_{k}+n_{-k})+\Delta _{k}^{\alpha }(-ic_{k}^{\dagger }c_{-k}^{\dagger }+%
\mathrm{{h.c.}),}
\label{H_k}
\end{equation}%
where $n_{k}:=c_{k}^{\dagger }c_{k}$ and the $c_{k}$'s are fermionic
operators i.e., $\{c_{k},c_{k^{\prime }}^{\dagger }\}=\delta _{k,{k^{\prime }%
}}.$ The Hilbert space factorizes $\mathcal{H}=\otimes _{k}\mathcal{H}_{k}.$
One has $\mathcal{H}_{k}\otimes \mathcal{H}_{-k}=\mathrm{{span}\{|00\rangle
_{k,-k},|11\rangle _{k,-k},|01\rangle _{k,-k},|10\rangle _{k,-k}\}.}$ The
first (last) two vectors span the even (odd) parity sector. The Hamiltonian (%
\ref{H_k}) has a trivial action over the odd sector i.e., $H_{k}^{\alpha
}|_{odd}=\epsilon _{k}^{\alpha }\openone_{k,-k}.$ Therefore, neglecting
irrelevant constants one can write
\begin{equation}
H_{k}^{\alpha }=2\epsilon _{k}^{\alpha }J_{k}^{z}+2\Delta _{k}^{\alpha
}J_{k}^{y},
\end{equation}%
where $J_{k}^{z}:=1/2(n_{k}+n_{-k}-1),\,J_{k}^{y}:=1/2\left(
-ic_{k}^{\dagger }c_{-k}^{\dagger }+\mathrm{h.c.}\right) \mathrm{.}$ For
every $k,$ the operators $J_{k}^{x}:=1/2(c_{k}^{\dagger }c_{-k}^{\dagger }+%
\mathrm{h.c.}),J_{k}^{y},J_{k}^{z}$ span a $su(2)$ Lie-algebra such that the
even (odd) sector of $\mathcal{H}_{k}\otimes \mathcal{H}_{-k}$ is the $J=1/2$
($J=0$) irreducible representation. Therefore, with obvious notation one can
write
\begin{eqnarray}
H_{k}^{\alpha } &=&\left( \epsilon _{k}^{\alpha }\sigma _{k}^{z}+\Delta
_{k}^{\alpha }\sigma _{k}^{y}\right) \oplus 0_{2}=\Lambda _{k}^{\alpha
}J_{k}^{\alpha }\oplus 0_{2}  \notag \\
&=&\left( e^{i\frac{\theta _{k}^{\alpha }}{2}\sigma _{kx}}\Lambda
_{k}^{\alpha }\sigma _{kz}e^{-i\frac{\theta _{k}^{\alpha }}{2}\sigma
_{kx}}\right) \oplus 0_{2}.
\end{eqnarray}%
where%
\begin{eqnarray*}
\Lambda _{k}^{\alpha } &:&=\sqrt{(\epsilon _{k}^{\alpha })^{2}+(\Delta
_{k}^{\alpha })^{2}},\, \\
J_{k}^{\alpha } &:&=\cos \theta _{k}^{\alpha }\sigma _{k}^{z}+\sin \theta
_{k}^{\alpha }\sigma _{k}^{y}, \\
\theta _{k}^{\alpha } &:&=\tan ^{-1}(\Delta _{k}^{\alpha }/\epsilon
_{k}^{\alpha }).
\end{eqnarray*}%
Moreover,
\begin{eqnarray}
\exp (-\beta _{\alpha }H_{k}^{\alpha }) &=&\varrho _{k}^{\alpha }(\beta
_{\alpha })\oplus \openone_{k,-k},  \notag \\
\rho _{\alpha } &=&\prod\limits_{k}\left( Z_{k}^{\alpha }\right)
^{-1}\varrho _{k}^{\alpha }(\beta _{\alpha })\oplus \openone_{k,-k}
\end{eqnarray}%
where
\begin{eqnarray}
\varrho _{k}^{\alpha }(\beta ) &:&=\exp (-\beta \Lambda _{k}^{\alpha
}J_{k}^{\alpha })  \notag \\
&=&\cosh (\beta \Lambda _{k}^{\alpha })-J_{k}^{\alpha }\sinh (\beta \Lambda
_{k}^{\alpha })  \label{qqq}
\end{eqnarray}%
is a $2\times 2$ operator in the even sector, and
\begin{equation}
Z_{k}^{\alpha }=2+2\cosh (\beta \Lambda _{k}^{\alpha }).  \label{zzz}
\end{equation}

The fidelity for the two thermal states is then given by
\begin{eqnarray}
&&\mathcal{F}_{H_{0},H_{1}}(\beta _{0},\beta _{1})  \notag \\
&=&\prod\limits_{k}\frac{2+\mathrm{{tr}\sqrt{\varrho _{k}^{1}(\beta _{1})^{%
\frac{1}{2}}\varrho _{k}^{0}(\beta _{0})\varrho _{k}^{1}(\beta _{1})^{\frac{1%
}{2}}}}}{\sqrt{Z_{k}^{0}Z_{k}^{1}}}.  \label{fff}
\end{eqnarray}%
So, apart for the trivial terms in the odd sector in order to compute the
fidelity, one has to consider the products $\varrho _{k}^{1}(\beta
_{1}/2)\varrho _{k}^{0}(\beta _{0})\varrho _{k}^{1}(\beta _{1}/2).$ As this
is a product of 2$\times 2$\quad matrices, one has
\begin{eqnarray*}
&&\mathrm{{tr}\sqrt{\varrho _{k}^{1}(\beta _{1})^{\frac{1}{2}}\varrho
_{k}^{0}(\beta _{0})\varrho _{k}^{1}(\beta _{1})^{\frac{1}{2}}}} \\
&\mathrm{=}&\sqrt{\text{Tr}\left[ \varrho _{k}^{0}(\beta _{0})\varrho
_{k}^{1}(\beta _{1})\right] +2\det (\varrho _{k}^{0}(\beta _{0})\varrho
_{k}^{1}(\beta _{1})}
\end{eqnarray*}%
Note that here $\det (\varrho _{k}^{0}(\beta _{0})\varrho _{k}^{1}(\beta
_{1})=1$, then substituting the above equation to Eq.(\ref{fff}) leads to a
simple expression of the fidelity
\begin{eqnarray}
&&\mathcal{F}_{H_{0},H_{1}}(\beta _{0},\beta _{1})  \notag \\
&=&\prod\limits_{k}\frac{2+\sqrt{\text{Tr}\left[ \varrho _{k}^{0}(\beta
_{0})\varrho _{k}^{1}(\beta _{1})\right] +2}}{\sqrt{Z_{k}^{0}Z_{k}^{1}}}.
\label{aaa}
\end{eqnarray}%
Now, we are left to compute the trace of $\varrho _{k}^{0}(\beta
_{0})\varrho _{k}^{1}(\beta _{1}).$

The matrix product $\varrho _{k}^{0}(\beta _{0})\varrho _{k}^{1}(\beta _{1})$
can be written as
\begin{eqnarray}
&&\text{Tr}\left[ \varrho _{k}^{0}(\beta _{0})\varrho _{k}^{1}(\beta _{1})%
\right]  \notag \\
&=&\text{Tr}\left( e^{-\beta _{0}\Lambda _{k}^{0}\sigma _{kz}}e^{i\alpha
_{k}\sigma _{kx}}e^{-\beta _{1}\Lambda _{k}^{1}\sigma _{kz}}e^{-i\alpha
_{k}\sigma _{kx}}\right)  \notag \\
&=&\text{Tr}(\left[ \cosh (\beta _{0}\Lambda _{k}^{0})-\sinh (\beta
_{0}\Lambda _{k}^{0})\sigma _{kz}\right]  \notag \\
&&\times e^{i\alpha _{k}\sigma _{kx}}e^{-\beta _{1}\Lambda _{k}^{1}\sigma
_{kz}}e^{-i\alpha _{k}\sigma _{kx}})  \notag \\
&=&2\cosh (\beta _{0}\Lambda _{k}^{0})\cosh (\beta _{1}\Lambda
_{k}^{1})-\sinh (\beta _{0}\Lambda _{k}^{0})  \notag \\
&&\times \text{Tr}\left( e^{-i\alpha _{k}\sigma _{kx}}\sigma _{kz}e^{-\beta
_{1}\Lambda _{k}^{1}\sigma _{kz}}e^{-i\alpha _{k}\sigma _{kx}}\right) ,
\label{var}
\end{eqnarray}%
where $\alpha _{k}=\frac{\theta _{k}^{1}-\theta _{k}^{0}}{2}.$

The following trace can be evaluated as
\begin{eqnarray}
&&\text{Tr}\left( e^{-i\alpha _{k}\sigma _{kx}}\sigma _{kz}e^{-\beta
_{1}\Lambda _{k}^{1}\sigma _{kz}}e^{-i\alpha _{k}\sigma _{kx}}\right)  \notag
\\
&=&\text{Tr}\left( e^{-i2\alpha _{k}\sigma _{kx}}\left[ \sigma _{kz}\cosh
(\beta _{1}\Lambda _{k}^{1})-\sinh (\beta _{1}\Lambda _{k}^{1})\right]
\right)  \notag \\
&=&\cosh (\beta _{1}\Lambda _{k}^{1})\text{Tr}\left( e^{-i2\alpha _{k}\sigma
_{kx}}\sigma _{kz}\right) -2\sinh (\beta _{1}\Lambda _{k}^{1})\cos (2\alpha
_{k})  \notag \\
&=&-2\sinh (\beta _{1}\Lambda _{k}^{1})\cos (2\alpha _{k})  \label{varvar}
\end{eqnarray}%
Substituting Eq.(\ref{varvar}) to (\ref{var}) leads to%
\begin{eqnarray*}
&&\text{Tr}\left[ \varrho _{k}^{0}(\beta _{0})\varrho _{k}^{1}(\beta _{1})%
\right] \\
&=&2[\cosh (\beta _{0}\Lambda _{k}^{0})\cosh (\beta _{1}\Lambda _{k}^{1}) \\
&&+\sinh (\beta _{0}\Lambda _{k}^{0})\sinh (\beta _{1}\Lambda _{k}^{1})\cos
(\theta _{k}^{0}-\theta _{k}^{1})].
\end{eqnarray*}%
Bringing all the terms together one eventually finds
\begin{eqnarray}
&&\mathcal{F}_{H_{0},H_{1}}(\beta _{0},\beta _{1})  \notag \\
&=&\prod_{k}\left\{ \left[ 1+\cosh (\beta _{0}\Lambda _{k}^{0})\right] \left[
1+\cosh (\beta _{1}\Lambda _{k}^{1})\right] \right\} ^{-\frac{1}{2}}  \notag
\\
&&\times (1+\frac{1}{\sqrt{2}}[1+\cosh (\beta _{0}\Lambda _{k}^{0})\cosh
(\beta _{1}\Lambda _{k}^{1})  \notag \\
&&+\sinh (\beta _{1}\Lambda _{k}^{0})\sinh (\beta _{1}\Lambda _{k}^{1})\cos
(\theta _{k}^{0}-\theta _{k}^{1})]^{\frac{1}{2}})  \label{Fid}
\end{eqnarray}%
It is easy to check that
\begin{eqnarray}
&&\mathcal{F}_{H_{0},H_{1}}(\beta ,\beta )\overset{\beta \rightarrow \infty }%
{\rightarrow }\prod_{k}\sqrt{\frac{1+\cos (\theta _{k}^{0}-\theta _{k}^{1})}{%
2}}  \notag \\
&=&\prod_{k}\left\vert \cos (\frac{\theta _{k}^{0}-\theta _{k}^{1}}{2}%
)\right\vert
\end{eqnarray}%
i.e., one recovers the zero-temperature result \cite{za-pa}.

\subsection{Numerical analysis}

We use XY model as an example to demonstrate our main idea. Here, two
Hamiltonians are $H_{0}=H(\gamma ,\lambda )$ and $H_{1}=H(\gamma +\delta
\gamma ,\lambda +\delta \lambda )$ where
\begin{equation}
H(\gamma ,\lambda )=\sum_{i}\left[ \frac{1+\gamma }{2}\sigma _{i}^{x}\sigma
_{i+1}^{x}+\frac{1-\gamma }{2}\sigma _{i}^{y}\sigma _{i+1}^{y}+\lambda
\sigma _{i}^{z}\right] ,  \label{xy}
\end{equation}%
where $\gamma $ defines the anisotropy and $\lambda $ represents external
magnetic field along the $z$ axis. Obviously, when $\gamma =1$, the XY
Hamiltonian (\ref{xy}) reduce to the transverse Ising Hamiltonian. $\delta
\gamma $ and $\delta \lambda $ represent small perturbation to the
Hamiltonian. $\sigma _{i}^{\alpha }$, $\alpha \in \{x,y,z\}$ are usual Pauli
operators in the $i$-th lattice point. It is well known that this model, by
means of a Jordan-Wigner transformation, can be mapped onto a quasi-free
fermionic Hamiltonian of the type (\ref{H_k}) [i.e., $\epsilon _{k}=\cos
\frac{2k\pi }{N}-\lambda ,\,\Delta _{k}=\gamma \sin \frac{2k\pi }{N}$].
Through a straightforward calculation, it can be obtained that $\Lambda
_{k}^{0}=\Lambda _{k}(\gamma ,\lambda )$ and $\Lambda _{k}^{1}=\Lambda
_{k}(\gamma +\delta \gamma ,\lambda +\delta \lambda )$ where
\begin{equation}
\Lambda _{k}^{0}=\sqrt{\left[ \cos \frac{2k\pi }{N}-\lambda \right]
^{2}+\gamma ^{2}\sin ^{2}\frac{2k\pi }{N}},
\end{equation}%
and $\theta _{k}^{\alpha }=\cos ^{-1}[(\cos \frac{2k\pi }{N}-\lambda
_{\alpha })/\Lambda _{k}^{\alpha }]$, $(\alpha =0,1$, $\lambda _{0}=\lambda
,\lambda _{1}=\lambda +\delta \lambda )$. 
We plot the mixed-state fidelity according to the analytical expression
obtained above (\ref{Fid}). We choose the number of spins to be $200$, and
the perturbation to be $10^{-2}$. In Fig. 1, we consider four different
temperatures; $\beta =1$, $\beta =10$, $\beta =20$, and $\beta =100$,
respectively. At low temperature, e.g., $\beta =100$, the sharp decay of
fidelity in critical region clearly displays the QPT, which happens at zero
temperature. What is more, the pattern of fidelity in low temperature is
very similar to the ground-state fidelity \cite{za-pa} of the XY model. This
similarity is natural because, as we mentioned above, the mixed-state
fidelity approaches ground-state fidelity when the temperature decreases to
zero. When the temperature increases higher, e.g., $\beta =10$ or $\beta =20$%
, the decay of fidelity, though less sharp due to the thermal excitation,
still early shows the signature of QPT. However, when the temperature
becomes large enough, e.g., $\beta =1$, the dramatic decay of ground-state
fidelity at critical point is totally washed out by the thermal excitation
and the signature of QPT at critical point eventually disappears. In Fig. 2,
we plot the mixed-state fidelity of thermal state of the transverse Ising
model. This is actually a cross section of Fig. 1 at $\gamma =1$. The decay
of fidelity clearly indicate the critical point of transverse Ising system.

It is important to stress that these numerical findings, besides
illustrating the possibility of detecting and studying the
finite-temperature signatures of QPTs, show that the ground-state fidelity
approach to criticality is endowed with some robustness against
\textquotedblleft perturbations" of the ground state. In this case, in fact
we have seen that mixing the ground state with excited eigenstates does not
destroy the peculiar behavior of fidelity in the neighborhood of the QPTs
i.e.,the fidelity drop.

\section{Loschmidt echo}

\begin{figure}[tbp]
\begin{center}
\includegraphics[bb=72 295 498 627, width=8cm, clip]{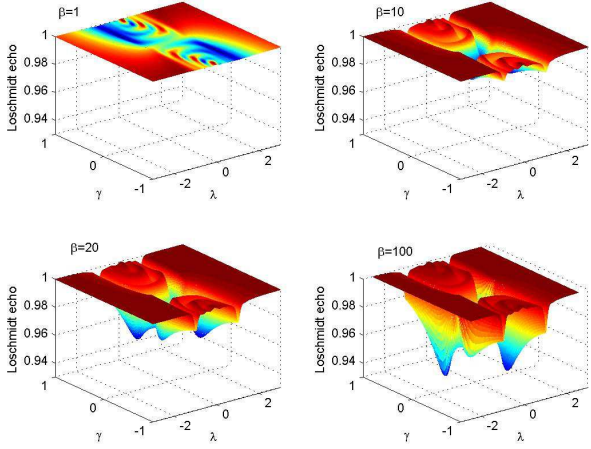}
\end{center}
\caption{(Color online) Mixed-state Loschmidt echo finite-temperature
thermal state of the XY model at an instant $t=10$. Here, we choose the spin
number $N=200$ and the parameter perturbation $\protect\delta \protect%
\lambda =\protect\delta \protect\gamma =10^{-2}$. The temperatures of system
in the above four figures are chosen to be $\protect\beta =1, \protect\beta %
=10, \protect\beta =20$, and $\protect\beta =100$, respectively. Similar to
the mixed-state fidelity approach, the critical region is well indicated by
Loschmidt echo at an instant when the system is at a low temperature.
However, the signature of QPT disappears when the temperature increases to
very high value.}
\end{figure}

\begin{figure}[tbp]
\begin{center}
\includegraphics[bb=36 190 568 608, width=7cm, clip]{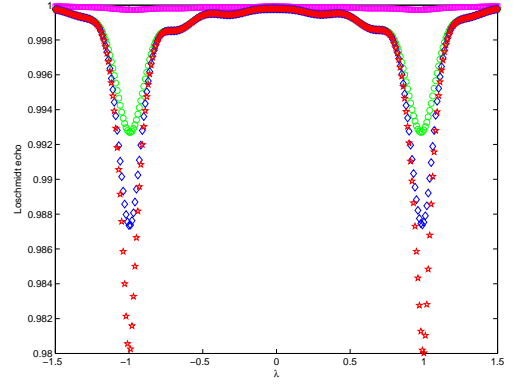}
\end{center}
\caption{(Color online) A cross section of Fig. 3 with $\protect\gamma=1$,
i.e., mixed-state Loschmidt echo (at an instant) of finite-temperature
thermal state of transverse Ising model. Here, the same as that in Fig. 3,
the spin number is chosen to be $N=200$ and the parameter perturbation $%
\protect\delta \protect\lambda =\protect\delta \protect\gamma =10^{-2}$.
Four curves represent four different temperatures: $\protect\beta =1,
\protect\beta =10, \protect\beta =20$, and $\protect\beta =100$,
respectively. The critical point is well indicated by Loschmidt echo when
the temperature is not too high.}
\end{figure}

In Refs. \cite{zhong-guo} and \cite{cuc}, the concept of Loschmidt echo has
been used to investigate quantum criticality. The resulting picture is that
the asymptotic value of the Loschmidt echo (for infinitely long times) is
directly related to the closeness of the system to the critical points i.e.,
the closer the systems to the QPT the smaller the asymptotic value of the
Loschmidt echo. On the other hand, even the short time behavior brings about
information about the QPTs. Indeed, for short times the decay of the
Loschmidt echo appears to be gaussian $\sim \exp (-\alpha t^{2}),$ where the
rate $\alpha $ has a diverging derivative (as a function of e.g., magnetic
field) at the critical point \cite{cuc}.

Now we show how to generalize the notion of Loschmidt echo in a natural way
to the thermal state case. This natural extension to the realm of
mixed-state can be obtained by using Eq. (\ref{mix-fid}),
\begin{eqnarray}
&&L_{\rho _{in},H_{0},H_{1}}(t):=F(\rho _{in},U_{1}(t)^{\dagger
}U_{0}(t)\rho _{in}U_{0}^{\dagger }(t)U_{1}(t)),  \notag \\
&&U_{\alpha }:=\exp (-iH_{\alpha }t)\quad (\alpha =0,1).  \label{Losch-echo}
\end{eqnarray}%
In particular, one can chose $\rho _{in}=\exp (-\beta H_{0}),$ then Eq. (\ref%
{Losch-echo}) simplifies and one can define
\begin{eqnarray}
\mathcal{L}_{H_{0},H_{1}}(\beta ,t) &=&F\left[ Z_{0}^{-1}e^{-\beta
H_{0}},Z_{0}^{-1}U_{1}^{\dagger }(t)e^{-\beta H_{0}}U_{1}(t)\right]  \notag
\\
&=&Z_{0}^{-1}F\left[ e^{-\beta H_{0}},U_{1}^{\dagger }(t)e^{-\beta
H_{0}}U_{1}(t)\right]  \notag \\
&=&\mathcal{F}_{H_{0},U_{1}^{\dagger }(t)H_{0}U_{1}(t)}(\beta ,\beta ).
\label{temp-echo}
\end{eqnarray}

Then, from the above equation and Eq. (\ref{aaa}), one finds
\begin{equation}
\mathcal{L}_{H_{0},H_{1}}(\beta ,t)=\prod\limits_{k}\left( Z_{k}^{0}\right)
^{-1}\left( 2+\sqrt{\text{Tr}\left[ \varrho _{k}^{0}(\beta )\varrho
_{k}^{1}(\beta )\right] +2}\right)
\end{equation}

with
\begin{eqnarray}
\varrho _{k}^{0}(\beta ) &=&e^{i\frac{\theta _{k}^{0}}{2}\sigma
_{kx}}e^{-\beta \Lambda _{k}^{0}\sigma _{kz}}e^{-i\frac{\theta _{k}^{0}}{2}%
\sigma _{kx}},  \notag \\
\varrho _{k}^{1}(\beta ) &=&U_{1}^{\dagger }(t)e^{i\frac{\theta _{k}^{0}}{2}%
\sigma _{kx}}e^{-\beta \Lambda _{k}^{0}\sigma _{kz}}e^{-i\frac{\theta
_{k}^{0}}{2}\sigma _{kx}}U_{1}(t).
\end{eqnarray}

Notice now that the unitary operator $U_1(t)$ can be written as
$U_{1}(t)=e^{i\frac{\theta _{k}^{1}}{2}\sigma _{kx}}e^{-i\Lambda
_{k}^{1}t\sigma _{kz}}e^{-i\frac{\theta _{k}^{1}}{2}\sigma _{kx}}. $
Then,
\begin{eqnarray}
&&\text{Tr}\left[ \varrho _{k}^{0}(\beta )\varrho _{k}^{1}(\beta )\right]
\notag \\
&=&\text{Tr(}e^{i\frac{\theta _{k}^{0}}{2}\sigma _{kx}}e^{-\beta \Lambda
_{k}^{0}\sigma _{kz}}e^{-i\frac{\theta _{k}^{0}}{2}\sigma _{kx}}e^{i\frac{%
\theta _{k}^{1}}{2}\sigma _{kx}}e^{it\Lambda _{k}^{1}\sigma _{kz}}e^{-i\frac{%
\theta _{k}^{1}}{2}\sigma _{kx}}  \notag \\
&&\times e^{i\frac{\theta _{k}^{0}}{2}\sigma _{kx}}e^{-\beta \Lambda
_{k}^{0}\sigma _{kz}}e^{-i\frac{\theta _{k}^{0}}{2}\sigma _{kx}}e^{i\frac{%
\theta _{k}^{1}}{2}\sigma _{kx}}e^{-it\Lambda _{k}^{1}\sigma _{kz}}e^{-i%
\frac{\theta _{k}^{1}}{2}\sigma _{kx}})  \notag \\
&=&\text{Tr(}e^{-\beta \Lambda _{k}^{0}\sigma _{kz}}e^{i\alpha _{k}\sigma
_{kx}}e^{it\Lambda _{k}^{1}\sigma _{kz}}e^{-i\alpha _{k}\sigma _{kx}}  \notag
\\
&&e^{-\beta \Lambda _{k}^{0}\sigma _{kz}}e^{i\alpha _{k}\sigma
_{kx}}e^{-it\Lambda _{k}^{1}\sigma _{kz}}e^{-i\alpha _{k}\sigma _{kx}})
\notag \\
&=&\text{Tr(}e^{-\beta \Lambda _{k}^{0}\sigma _{kz}}e^{-\beta \Lambda
_{k}^{0}V^{\dagger }\sigma _{kz}V}),
\end{eqnarray}%
where $V=e^{i\alpha _{k}\sigma _{kx}}e^{-it\Lambda _{k}^{1}\sigma
_{kz}}e^{-i\alpha _{k}\sigma _{kx}}.$ Using the formula $e^{x}=\cosh x-\sinh
x,$ one have
\begin{eqnarray}
&&\text{Tr}\left[ \varrho _{k}^{0}(\beta )\varrho _{k}^{1}(\beta )\right]
\notag \\
&=&2\cosh ^{2}(\beta \Lambda _{k}^{0})+\sinh ^{2}(\beta \Lambda _{k}^{0})%
\text{Tr(}\sigma _{kz}V^{\dagger }\sigma _{kz}V).  \label{bbb}
\end{eqnarray}

The following trace is evaluated as
\begin{eqnarray}
&&\text{Tr(}\sigma _{kz}V^{\dagger }\sigma _{kz}V)  \notag \\
&=&\text{Tr(}\sigma _{kz}e^{i\alpha _{k}\sigma _{kx}}e^{it\Lambda
_{k}^{1}\sigma _{kz}}e^{-i\alpha _{k}\sigma _{kx}}  \notag \\
&&\times \sigma _{kz}e^{i\alpha _{k}\sigma _{kx}}e^{-it\Lambda
_{k}^{1}\sigma _{kz}}e^{-i\alpha _{k}\sigma _{kx}})  \notag \\
&=&\text{Tr(}\sigma _{kz}e^{i\alpha _{k}\sigma _{kx}}e^{it\Lambda
_{k}^{1}\sigma _{kz}}e^{-i2\alpha _{k}\sigma _{kx}}e^{-it\Lambda
_{k}^{1}\sigma _{kz}}e^{i\alpha _{k}\sigma _{kx}}\sigma _{kz})  \notag \\
&=&\text{Tr(}e^{i2\alpha _{k}\sigma _{kx}}e^{it\Lambda _{k}^{1}\sigma
_{kz}}e^{-i2\alpha _{k}\sigma _{kx}}e^{-it\Lambda _{k}^{1}\sigma _{kz}})
\notag \\
&=&\text{Tr(}\left[ \cos \text{(}t\Lambda _{k}^{1}\text{)+}i\sin \text{(}%
\Lambda _{k}^{1}t\text{)}e^{i2\alpha _{k}\sigma _{kx}}\sigma
_{kz}e^{-i2\alpha _{k}\sigma _{kx}}\right]  \notag \\
&&\times \left[ \text{cos(}t\Lambda _{k}^{1}\text{)-}i\text{sin(}t\Lambda
_{k}^{1}\text{)}\sigma _{kz}\right] )  \notag \\
&=&2\text{cos}^{2}\text{(}t\Lambda _{k}^{1}\text{)+sin}^{2}\text{(}t\Lambda
_{k}^{1}\text{)Tr(}\sigma _{kz}e^{i2\alpha _{k}\sigma _{kx}}\sigma
_{kz}e^{-i2\alpha _{k}\sigma _{kx}})  \notag \\
&=&2\text{cos}^{2}\text{(}t\Lambda _{k}^{1}\text{)+}2\text{sin}^{2}\text{(}%
t\Lambda _{k}^{1}\text{)cos(}4\alpha _{k}\text{)}
\end{eqnarray}

Substituting the above equation to Eq.(\ref{bbb}) leads to
\begin{eqnarray}
&&\text{Tr}\left[ \varrho _{k}^{0}(\beta )\varrho _{k}^{1}(\beta )\right]
\notag \\
&=&2\cosh ^{2}(\beta \Lambda _{k}^{0})+2\sinh ^{2}(\beta \Lambda _{k}^{0})
\notag \\
&&\times \left[ \text{cos}^{2}\text{(}t\Lambda _{k}^{1}\text{)+sin}^{2}\text{%
(}t\Lambda _{k}^{1}\text{)cos(}4\alpha _{k}\text{)}\right]  \notag \\
&=&2\left[ 1-\sin ^{2}(2\alpha _{k})\sin ^{2}(\Lambda _{k}^{1}t)\right]
\cosh (2\beta \Lambda _{k}^{0})  \notag \\
&&+2\sin ^{2}(2\alpha _{k})\sin ^{2}(\Lambda _{k}^{1}t).
\end{eqnarray}

Bringing all the terms together, we give an explicit expression for the
Loschmidt echo for the class of Hamiltonians (\ref{H_k})
\begin{eqnarray}
&&\mathcal{L}_{H_{0},H_{1}}(\beta ,t)  \label{echoecho} \\
&=&\prod_{k}\left[ 1+\cosh (\beta \Lambda _{k}^{0})\right] ^{-1}  \notag \\
&&\times \{1+\frac{1}{\sqrt{2}}\{\left[ 1-\sin ^{2}(\theta _{k}^{0}-\theta
_{k}^{1})\sin ^{2}(\Lambda _{k}^{1}t)\right]   \notag \\
&&\times \cosh (2\beta \Lambda _{k}^{0})+\sin ^{2}(\theta _{k}^{0}-\theta
_{k}^{1})\sin ^{2}(\Lambda _{k}^{1}t)+1\}^{1/2}\}  \notag
\end{eqnarray}

It is immediate to check that in the zero temperature limit one recovers the
result obtained in~\cite{zhong-guo},
\begin{equation*}
\mathcal{L}_{H_{0},H_{1}}(\beta ,t)\overset{\beta \rightarrow \infty }{%
\rightarrow }\prod_{k}\sqrt{(1-\sin ^{2}(\theta _{k}^{0}-\theta
_{k}^{1})\sin ^{2}(\Lambda _{k}^{1}t)}.
\end{equation*}%
%
%
%
%
%

\subsection{Numerical analysis}

Similarly to Sec. II, we now study mixed-state Loschmidt echo given by Eq. (%
\ref{echoecho}). In Fig. 3, we plot the Loschmidt echo at an instant $t=10$.
The parameters are the same as that in Fig. 1 and Fig. 2. Similar to the
ground-state Loschmidt echo in Ref. \cite{zhong-guo}, the decay of
mixed-state Loschmidet echo of a thermal state at a finite temperature well
indicates the critical point. With the increase of the temperature, the
decay of the mixed-state Loschmidt echo at critical point becomes less sharp
until finally disappears. Fig. 4 is a cross section of Fig. 3 at $\gamma =1$%
, i.e., a mixed-state Loschmidt echo of transverse Ising model. Similarly to
the mixed-state Fidelity of transverse Ising model in Fig. 2, the critical
point is clearly indicated by the Loschmidt echo. With temperature increase,
the decay of Loschmidt echo at critical point becomes less evident, and
finally disappears for sufficiently high temperature.

\section{Conclusions}

In this paper we have shown how to extend to finite temperature the fidelity
approach to quantum phase transitions advocated in References \cite{za-pa}
and \cite{za-co-gio}. This generalization relies on the notion of mixed
state fidelity applied to (Gibbs) thermal states. Mixed-state fidelity is
strictly related to the Bures metric measuring the statistical distance
between two density operators, therefore this approach has a geometrical as
well as an operational meaning and two of them are deeply intertwined. We
provided an explicit analytical expression for both the fidelity and
Loschmidt echo for an important class of quasi-free fermionic Hamiltonians
including e.g., the XY model. A numerical analysis of these quantities has
been performed; it clearly shows how the influence of the zero-temperature
critical points extend over a finite range of temperatures. Besides
representing a non-trivial generalization of the former zero-temperature
results, the findings reported in this paper suggest directions for further
investigations. The most prominent being the possibility of using (mixed)
state-space metrical quantities to study temperature driven phase
transitions both in the quantum and in the classical case \cite{gio}. The
fact that simple geometrical notions seme to provide a unified frame for
studying all these different kinds of critical phenomena is, we believe,
conceptually quite appealing.

\acknowledgements The authors thank for the discussions with Y. Q. Li. XW is
supported by CNSF under grant No. 10405019, Specialized Research Fund for
the Doctoral Program of Higher Education (SRFDP) under grant No.20050335087,
and The Project-sponsored by SRF for ROCS, SEM. CPS is supported by the NSFC
with grant Nos. 90203018, 10474104 and 60433050, and the National
Fundamental Research Program of China with Nos. 2001CB309310 and
2005CB724508.

\end{document}